\documentclass{caosp}

\usepackage{graphicx}
\usepackage[utf8]{inputenc}
\usepackage[T1]{fontenc}

\articleNo{0}
\pubyear{0}
\volume{0}
\volnumber{0}
\firstpage{1}
\received{}
\accepted{}

\begin{document}

\title{The hot spot in HT Cas during its Superoutburst}

\author{
	K.\,B\k{a}kowska \inst{1,} \inst{2}
		\and
		A.\,Olech \inst{1}
}

\institute{
		Nicolaus Copernicus Astronomical Center, Polish Academy of Sciences\\
		ul. Bartycka 18, 00-716 Warszawa, Poland \\
		\and
		Astronomical Observatory Institute, Faculty of Physics,\\
		A. Mickiewicz University,
		ul. S{\l}oneczna 36, 60-286 Pozna\'{n}, Poland\\
}

\date{October 7, 2013}

\maketitle

\begin{abstract}
We present the analysis of eclipses observed in dwarf nova HT Cas during its superoutburst in November 2010. Detection of hot spot was confirmed. 
\keywords{Stars: individual: HT Cas - binaries: 
close - novae, cataclysmic variables}
\end{abstract}

\section{Introduction}

\label{intr}

Cataclysmic variables are close binary stars containing a white dwarf (the primary) and a main-sequence star (the secondary). In non-magnetic systems the material flows from the secondary and forms a luminous accretion disk around the primary. The hot spot is the location where the matter hits the accretion disk (for revision Hellier, 2001). 

In 2010 after 25 years of quiescence and normal outbursts HT Cas went into superoutburst. We present analysis of two eclipses observed during this unique phenomenon.  Reconstruction of the hot spot was made with the decomposition method. Details and beautiful examples of application of this method for cataclysmic variables UX UMa, Z Cha and OY Car were presented by Smak (1994\,a, 1994\,b, 2007, 2008\,b).

\section{Results}

\begin{figure}
\centerline{\includegraphics[width=5.5cm,clip=]{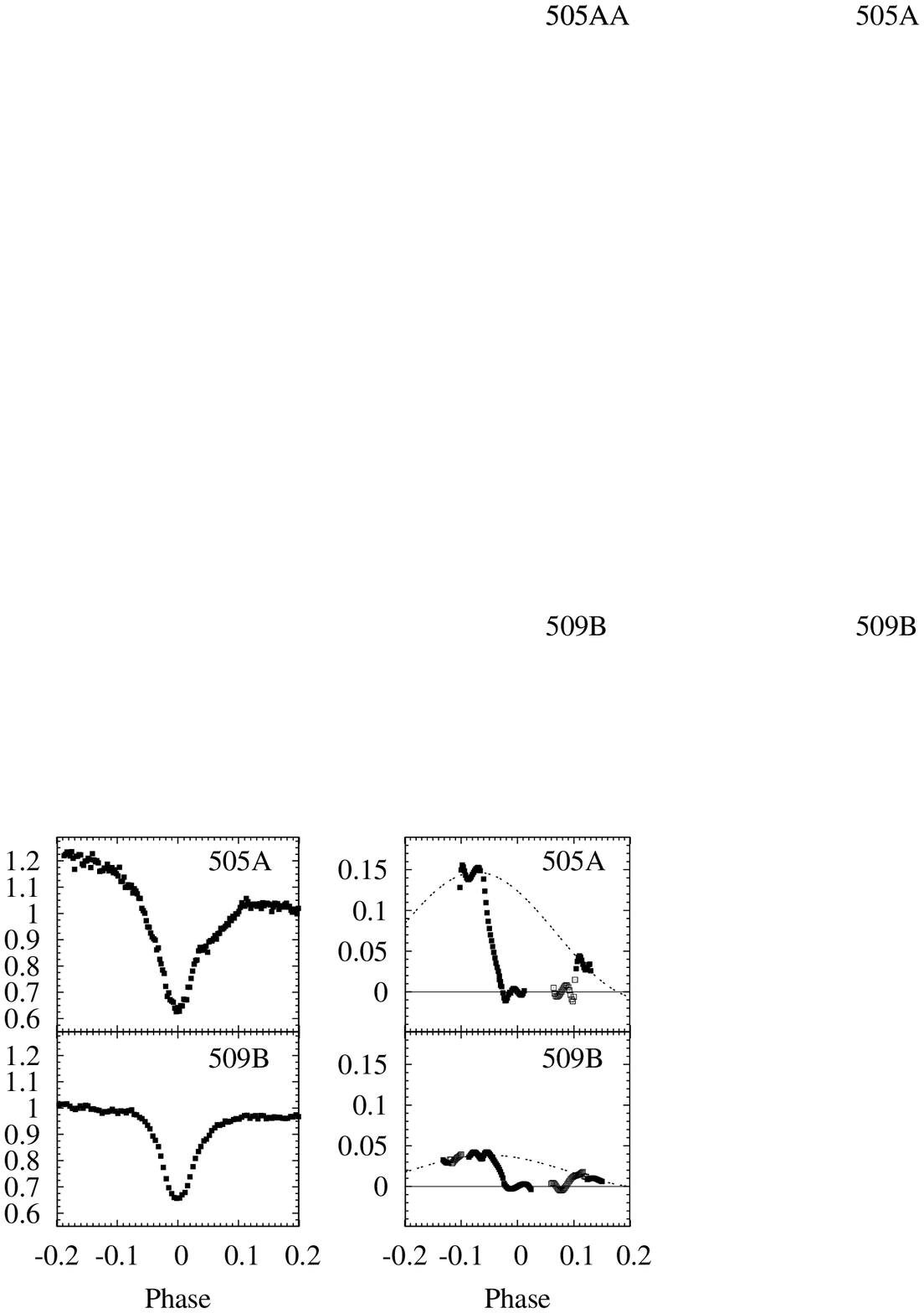}}
\caption{Left panel: observational data, on top the eclipse from the 2nd night of superoutburst (505 HJD), bottom panel: the eclipse from the 6th night of superoutburst (509 HJD). Right panel: reconstruction of hot spot obtained with decomposition method.}
\label{f1}
\end{figure}

The superoutburst in HT Cas from November 2010 lasted 11 nights (2455504 - 2455514 HJD, below marked as 504 - 514 HJD). In Fig.\,\ref{f1} on the left panel there are two light curves with eclipses observed in HT Cas. On the left top panel around phases $-0.1<\phi<-0.05$ and $0.05<\phi<0.1$ one can see the beginning and the end of the eclipse of the hot spot, and this fact is confirmed by the reconstruction of the  hot spot presented on the top right panel. On the bottom left panel we present the eclipse where the contribution of the hot spot is not so obvious to observe and the reconstruction presented on right bottom panel confirms that the luminosity of the hot spot is indeed noticeably smaller. 

The manifestation of the hot spot was explicit during second night of superoutburst (505 HJD). The amplitude of the hot spot luminosity was $l_{s,max}=14.7\%$ and the phase corresponding to the ampitude was $\phi_{max}=-0.072$. Four nights later on, the hot spot luminosity was clearly smaller with the amplitude of $l_{s,max}=3.85\%$ and the phase $\phi_{max}=-0.052$. 

Manifestation of the hot spot in HT Cas during its superoutburst confirms that mass transfer is strongly enchanced. It is in full agreement with EMT (Enchanced Mass Transfer) model proposed by Smak (1991, 2004, 2008\,a). Analysis made with decomposition method of eclipses observed during the November 2010 superoutburst in HT Cas gives the newest observational evidence for EMT model.  

\acknowledgements
We want to thank Prof. Smak for his thoughtful comments about the decomposition method. Project was supported by  Polish National Science Center grants awarded by decision DEC-2012/07/N/ST9/04172 for KB.

\end{document}